\documentclass[aps,12pt]{revtex4}
\usepackage{graphicx}
\usepackage{amsmath}
\usepackage{amssymb}
\usepackage{epsfig}

\def\be{\begin{equation}}
\def\ee{\end{equation}}
\def\bea{\begin{eqnarray}}
\def\eea{\end{eqnarray}}

\begin{document}
\baselineskip=16pt
\begin{titlepage}
\setcounter{page}{0}
\begin{center}
\vspace{0.5cm}
 {\Large \bf Cosmological Evolution of a Tachyon-Quintom Model of Dark Energy}\\
\vspace{10mm} Shang-Gang Shi$^{1}$, Yun-Song Piao$^{1}$ ~and~ Cong-Feng Qiao$^{1,2}$\\
\vspace{6mm} {\footnotesize{\it $^1$College of Physical
Sciences,Graduate University of CAS, YuQuan Road 19A,
Beijing 100049, China\\
$^2$Theoretical Physics Center for Science Facilities (TPCSF), CAS.,
Beijing 100049, China\\}} \vspace*{5mm} \normalsize
\smallskip
\medskip
\smallskip
\end{center}
\vskip0.6in \centerline{\large\bf Abstract} \vskip0.3in
{\hspace{0.6cm}In this work we study the cosmological evolution of a
dark energy model with two scalar fields, i.e. the tachyon and the
phantom tachyon. This model enables the equation of state $w$ to
change from $w>-1$ to $w<-1$ in the evolution of the universe. The
phase-space analysis for such a system with inverse square
potentials shows that there exists a unique stable critical point,
which has power-law solutions. In this paper, we also study another
form of tachyon-quintom model with two fields, which voluntarily
involves the interactions between both fields.}

\vspace*{2mm}
\end{titlepage}

\section{Introduction}
Recent observational data ~\cite{PR97,S03,R04} strongly indicate
that the Universe is spatially flat and accelerating at the present
time. Within the framework of general relativity, cosmic
acceleration can be sourced by an energy-momentum tensor which has a
large negative pressure called dark energy (See Ref.~\cite{CLW06}
for a recent review). The simplest candidate for dark energy seems
to be a small positive cosmological constant, but it suffers from
difficulties associated with the fine tuning and coincidence
problem. This problem can be alleviated in models of dynamically
evolving dark energy called quintessence ~\cite{Zlatev 99}, which
have tracker like properties where the energy density in the fields
track those of the background energy density before dominating
today. The phantom, whose kinetic energy term has the reverse sign,
has been also proposed as a candidate of dynamical dark
energy~\cite{Caldwell02}. There has been the enormous variety of DE
models suggested in the literature, see Ref. \cite{SS00} for
reviews.

The analysis of the properties of dark energy from recent
observations mildly favor models with $\omega$ crossing $-1$ in
the near past~\cite{Alam04,FWZ05}. But, neither quintessence nor
phantom can fulfill this transition. The quintom scenario of dark
energy is designed to understand the nature of dark energy with
$\omega$ across $-1$. The first model of quintom scenario of dark
energy is given in Ref. ~\cite{FWZ05} with two scalar fields,
where one is quintessence and the other is phantom. This model has
been studied in detail later on
\cite{FLPZ06,RGCai,XinZhang,AKV,Setare,Chimento}, and recently a
new type of quintom model inspired by the string theory has also
been proposed, which only have a single scalar field~\cite{Yi-Fu
Cai08}.

The role of the rolling tachyon~\cite{Sen02} in string theory has
been widely studied in cosmology, see Refs.\cite{Gibbons02,FT2002},
and especially Refs.\cite{P02,CGJP04,AL04,CGST04} for dark energy.
Some sort of tachyon condensate may described by effective field
theory with a Lagrangian density
$\mathcal{L}=-V(\phi)\sqrt{1+g^{\mu\nu}\partial_\mu\phi\partial_\nu\phi}$.
It can act as a source of dark matter or inflation field. Meanwhile
the tachyon can also act as a source of dark energy depending upon
the form of the tachyon potential. However, as compared to canonical
quintessence, tachyon models require more fine-tuning to agree with
observations. In Ref.~\cite{HL03}, the authors consider the
Born-Infeld type Lagrangian with negative kinetic energy term. The
Lagrangian density they choose is
$\mathcal{L}=-V(\varphi)\sqrt{1-g^{\mu\nu}\partial_\mu\varphi\partial_\nu\varphi}$.
It is clear that for the spatially homogeneous scalar field, the
equation of state $\omega=-1-\dot{\varphi}^2$ will be less than $-1$
unless the kinetic energy term $\dot{\varphi}^2= 0$. This field is
called phantom tachyon, see for example~\cite{TS04}.

In principle, we can consider the multi-fields model including
multiple tachyon and multiple phantom tachyon. However, without loss
of generality, we only consider the case of one tachyon and one
phantom tachyon, since it is the simplest one to realize that the
equation of state $w$ cross $-1$ during the evolution of the
universe and it shows most of the central ideas of such model.
Because of the quintom-like behavior it shows, we call it
tachyon-quintom for the convenience. This paper is organized as
follows: in section II we study in detail the tachyon-quintom model
with inverse square potentials. The numerical analysis shows this
model is not sensitive to the initial kinetic energy density of
tachyon and phantom tachyon, and we give the reason in detail. Then
the phase-space analysis of the spatially flat FRW models shows that
there exist a unique stable critical point, and we compare it with
tachyon model; in section III we present another two-field model
which include the interaction between two fields; the section IV is
summary.

\section{The Tachyon-Quintom Model}

\subsection*{A.~The tachyon-quintom model }
We assume a four-dimensional, spatially-flat
Friedmann-Robertson-Walker Universe filled by a homogeneous tachyon
$\phi$ with potential $V(\phi)$,  a homogeneous phantom tachyon
$\varphi$ with potential $V(\varphi)$ and a fluid with barotropic
equation of state $p_\gamma=(\gamma-1)\rho_\gamma$, $0<\gamma\leq2$,
such as radiation ($\gamma=4/3$)or dust matter ($\gamma=1$). In this
section, we turn our attention to the possibility of the tachyon and
phantom tachyon as a source of the dark energy.

The action for such a system is given by
\begin{equation}
 S= \int
d^4x\sqrt{-g} \left( {\frac {M_p^2\mathcal{R}}{2}+\mathcal{L}_\phi
+\mathcal{L}_\varphi+\mathcal{L}_m}\right)
\end{equation}

where $M_p$ is the reduced Planck mass, $\mathcal{R}$ is the scalar
curvature, $\mathcal{L}_m$ represents the Lagrangian density of
matter fields and
\begin{equation}
\label{Lagrangian density1}
\mathcal{L}_\phi=-V(\phi)\sqrt{1+g^{\mu\nu}\partial_\mu\phi\partial_\nu\phi}
\end{equation}
\begin{equation}
\label{Lagrangian density2}
\mathcal{L}_\varphi=-V(\varphi)\sqrt{1-g^{\mu\nu}\partial_\mu\varphi\partial_\nu\varphi}.
\end{equation}

We now restrict to spatially homogeneous time dependent solutions
for which $\partial_i\phi =\partial_i\varphi = 0$. Thus the energy
densities and the pressure of the field $\phi$ and $\varphi$ are
given, respectively, by
\begin{eqnarray}\label{e1}\rho_\phi=\frac{V(\phi)}{\sqrt{1-\dot{\phi}^2
 }},~~~~p_\phi=-V(\phi)\sqrt{1-\dot{\phi}^2 }
,
\\
\label{e2}\rho_\varphi=\frac{V(\varphi)}{\sqrt{1+\dot{\varphi}^2 }},
~~~~p_\varphi=-V(\varphi)\sqrt{1+\dot{\varphi}^2 }
\end{eqnarray}

Here a dot is derivation with respect to synchronous time. The
background equations of motion are
\begin{equation}
\label{eq1}
\frac{\ddot{\phi}}{1-\dot{\phi}^2}+3H\dot{\phi}+\frac{1}{V(\phi)}\frac{dV(\phi)}{d\phi}=0
\end{equation}

\begin{equation}
\label{eq2}\frac{\ddot{\varphi}}{1+\dot{\varphi}^2}+3H\dot{\varphi}-\frac{1}{V(\varphi)}\frac{dV(\varphi)}{d\varphi}=0
\end{equation}

\begin{equation}\dot{\rho}_\gamma=-3\gamma H\rho_\gamma
\end{equation}

\begin{equation}\begin{array}{l}\label{Hubble}
\dot{H}=-\frac{1}{2M_p^2}(\rho_\phi+p_\phi+\rho_\varphi+p_\varphi+\rho_\gamma+p_\gamma)\\~~
  =-\frac{1}{2M_p^2} \left(
{\frac{\dot{\phi}^2V(\phi)}{\sqrt{1-\dot{\phi}^2}}-\frac{\dot{\varphi}^2V(\varphi)}{\sqrt{1+\dot{\varphi}^2}}+\gamma\rho_\gamma}
\right)\end{array}\\
 \end{equation}

together with a constraint equation for the Hubble parameter:
\begin{equation}
\label{Hubbleeq}
 H^2=\frac{1}{3M_p^2}\left(
{\frac{V(\phi)}{\sqrt{1-\dot{\phi}^2}}+\frac{V(\varphi)}{\sqrt{1+\dot{\varphi}^2}}+
\rho_\gamma} \right)
\end{equation}

The potentials we considered are inverse square potentials:
\begin{equation}V(\phi)=M^2_\phi\phi^{-2},~~~V(\varphi)=M^2_\varphi\varphi^{-2}
\end{equation}
Those potentials allow constructing a autonomous system
~\cite{CLW97,HW02} using the evolution equations, and give power-law
solutions. The cosmological dynamics of the tachyon field with
inverse square potential was studied in Ref.~\cite{P02,AL04,CGST04}.
Interestingly, the inverse square potential plays the same role for
tachyon fields as the exponential potential does for standard scalar
fields.

We define the following dimensionless quantities :
\begin{equation}x_\phi\equiv\dot{\phi},~~y_\phi\equiv\frac{V(\phi)}{3H^2M_p^2},
~~x_\varphi\equiv\dot{\varphi},~~y_\varphi\equiv\frac{V(\varphi)}{3H^2M_p^2},~~z\equiv\frac{\rho_\gamma}{3H^2M_p^2}
\end{equation}
Now the Eqs.~(\ref{Hubbleeq}) and (\ref{Hubble}) can be rewrite as
follow:
\begin{equation}
\label{Hubble1}
 1=\frac{y_\phi}{\sqrt{1-x^2_\phi}}+\frac{y_\varphi}{\sqrt{1+x^2_\varphi}}+z\equiv\Omega_{DE}+z
\end{equation}
\begin{equation}
\label{Hubbleeq1}
  \frac{H'}{H}=-\frac{3}{2}\left( {-\frac{y_\phi (\gamma-x^2_\phi)}{\sqrt{1-x^2_\phi}}-\frac{y_\varphi (\gamma+x^2_\varphi)}{\sqrt{1+x^2_\varphi}}+\gamma} \right)
\end{equation}
where $\Omega_{DE}$ measure the dark energy density as a fraction of
the critical density,~a prime denotes a derivative with respect to
the logarithm of the scale factor,$N={\rm ln}\,a$.

Then the evolution Eqs.~(\ref{eq1}) and (\ref{eq2}) can be written
to an autonomous system:
\begin{equation}
\label{yundong1}
 x'_\phi=-3(x_\phi-\sqrt{\beta_\phi y_\phi })(1-x_\phi^2)
\end{equation}
\begin{equation}
\label{yundong2} y'_\phi=3 y_\phi\left( {-\frac{y_\phi
(\gamma-x^2_\phi)}{\sqrt{1-x^2_\phi}}-\frac{y_\varphi
(\gamma+x^2_\varphi)}{\sqrt{1+x^2_\varphi}}-\sqrt{\beta_\phi y_\phi
}x_\phi+\gamma} \right)
\end{equation}
\begin{equation}
\label{yundong3} x'_\varphi=-3(x_\varphi+\sqrt{\beta_\varphi
y_\varphi })(1+x_\varphi^2)
\end{equation}
\begin{equation}
\label{yundong4} y'_\varphi=3y_\varphi\left( {-\frac{y_\phi
(\gamma-x^2_\phi)}{\sqrt{1-x^2_\phi}}-\frac{y_\varphi
(\gamma+x^2_\varphi)}{\sqrt{1+x^2_\varphi}}-\sqrt{\beta_\varphi
y_\varphi }x_\varphi+\gamma} \right)
\end{equation}
where
\begin{equation}
 \beta_\phi=\frac{4M_p^2}{3M_\phi^2},~~~\beta_\varphi=\frac{4M_p^2}{3M_\varphi^2}
\end{equation}
 The equation of state of the dark energy is
\begin{equation}\label{w}
\omega=\frac{-V(\phi)\sqrt{1-\dot{\phi}^2}-V(\varphi)\sqrt{1+\dot{\varphi}^2}}{\frac{V(\phi)}{\sqrt{1-\dot{\phi}^2}}+\frac{V(\varphi)}{\sqrt{1+\dot{\varphi}^2}}}
=\frac{-y_\phi\sqrt{1-x_\phi^2}-y_\varphi\sqrt{1+x_\varphi^2}}{\frac{y_\phi}{\sqrt{1-x^2_\phi}}+\frac{y_\varphi}{\sqrt{1+x^2_\varphi}}}
\end{equation}

\subsection*{B.~Numerical analysis}
\begin{figure}
\centering
\includegraphics[height=2.5in,width=3.2in]{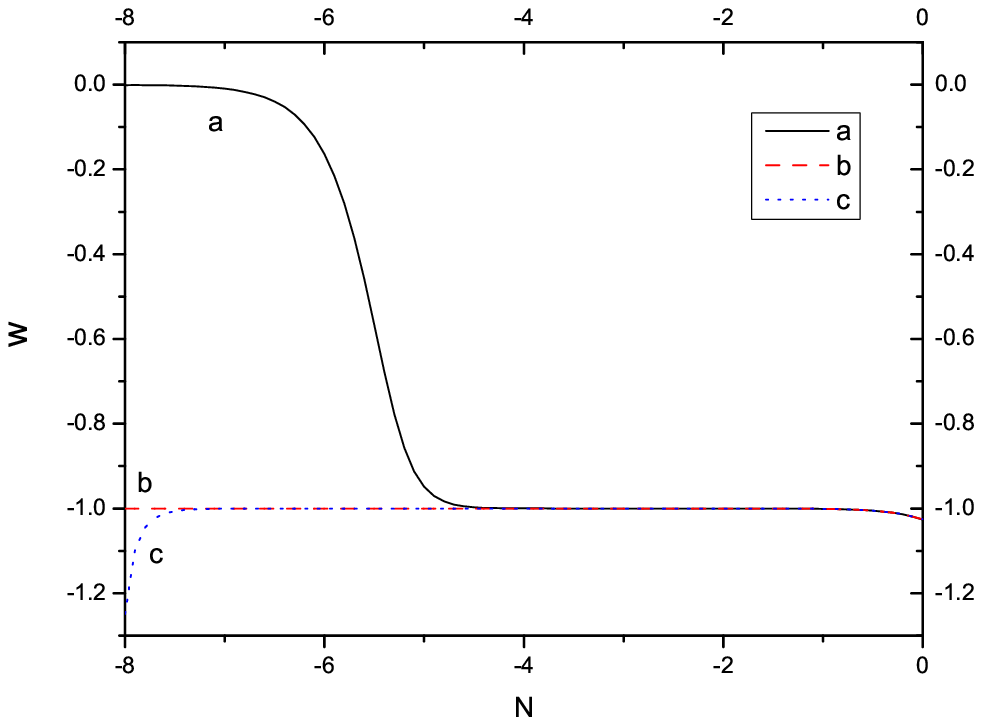}
\includegraphics[height=2.5in,width=3.2in]{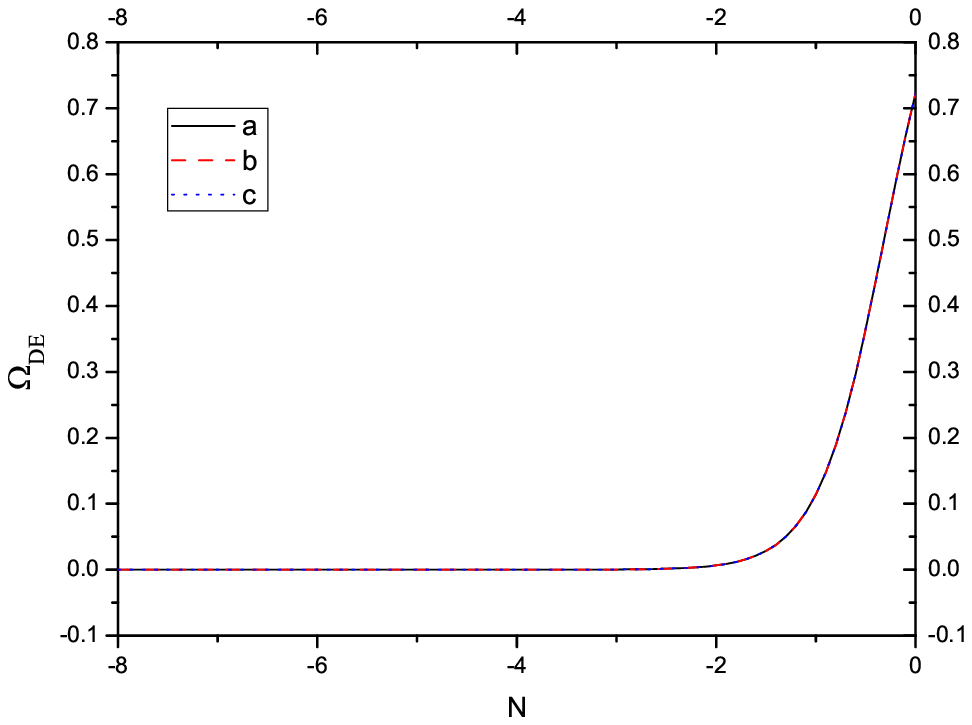}
\caption{\label{figure1} Evolution of the equation of state
($\omega$)and density parameters($\Omega_{DE}$) as a function of N
for the dark energy model with
$\beta_\phi=\beta_\varphi=1/3,~\gamma=1$.~Initial conditions~(at
$N=-8$) :~~a.~solid line:~$x_{\phi i}=0.9999999$,~$y_{\phi
i}=6\times10^{-11}$,~$x_{\varphi i}=2.0$,~$y_{\varphi
i}=6.5\times10^{-11}$;~~b.~dashed line:~$x_{\phi
i}=1\times10^{-12}$,~$y_{\phi i}=6\times10^{-11}$,~$x_{\varphi
i}=1\times10^{-12}$,~$y_{\varphi i}=6.5\times10^{-11}$;~~c.~dotted
line:~$x_{\phi i}=0.5$,~$y_{\phi i}=6\times10^{-11}$,~$x_{\varphi
i}=1.0$,~$y_{\varphi i}=6.5 \times10^{-11}$.}
\end{figure}

Mapping between the number of $e$-foldings and the redshift $z\equiv
a_0/a-1=1/a-1$,  we note that at big bang nucleosynthesis (BBN)
$N_{\rm BBN}\approx-20$ ($z\approx 10^9$), at matter-radiation
equality $N_{\rm eq}\approx -8$ ($z\approx 3200$). We choose $N=-8$
as the initial number of e-folds, so choose the $\gamma=1$ in
Eqs.~(\ref{yundong2}) ~and (\ref{yundong4}) is a good approximation.
The evolutions of $\omega$ and $\Omega_{DE}$ are shown in
Fig.\,\ref{figure1}. In Ref.~\cite{BHM01}, the authors use standard
Big Bang Nucleosynthesis and the observed abundances of primordial
nuclides to give a constraints on the scalar matter :
$\Omega_{DE}<0.045$, at temperatures near 1 MeV. The initial values
of $x_\phi,y_\phi,x_\varphi$~and ~$y_\varphi$ given below are safely
satisfy this requirement, since $\Omega_{DE}\leq 6\times10^{-7}$ at
$N=-8$, and the energy density of $\phi$ and $\varphi$ are
decreasing more slowly than the
fluid$(\omega_\phi=\frac{p_\phi}{\rho_\phi}=-1+\dot{\phi}^2<0,
\omega_\varphi=\frac{p_\varphi}{\rho_\varphi}=-1-\dot{\varphi}^2\leq-1
)$. So at temperatures near 1 MeV, the $\Omega_{DE}$ is smaller.
%%%%%%%%%%
%%%%%%%%%%

From pictures we can see that this model is not sensitive to the
initial kinetic energy density of tachyon and phantom
tachyon~($x_\phi=\dot{\phi},x_\varphi=\dot{\varphi}$). When $x_{\phi
i}=0.9999999,~y_{\phi i}=6\times10^{-11},~x_{\varphi
i}=2.0,~y_{\varphi i}=6.5\times10^{-11}$,
 the
 current $\omega$ and $\Omega_{DE}$ are $-1.025896$ , $0.722306$, respectively. When
$x_{\phi i}=1\times10^{-12},~y_{\phi i}=6\times10^{-11},~x_{\varphi
i}=1\times10^{-12},~y_{\varphi i}=6.5\times10^{-11}$, the
 current $\omega$ and $\Omega_{DE}$ are $-1.025841$ , $0.722367$, respectively.
 From Eq.~(\ref{e1}), we know that the initial energy density of the tachyon
varied by nearly four orders of magnitude is still consistent with
current observational constraints. But the initial potential energy
density of tachyon and phantom tachyon require fine-tuning to agree
with observations.

At the present, we want to explain in rough detail how solutions
converge to the common solution for the different initial conditions
which given in Fig.\,\ref{figure1}. (In next subsection, we will
know that there is only one stable critical point, so the solutions
converge to a common, cosmic evolutionary track is not surprised. )

%%%%%%%%%%
\begin{figure}
\centering
\includegraphics[height=2.5in,width=3.2in]{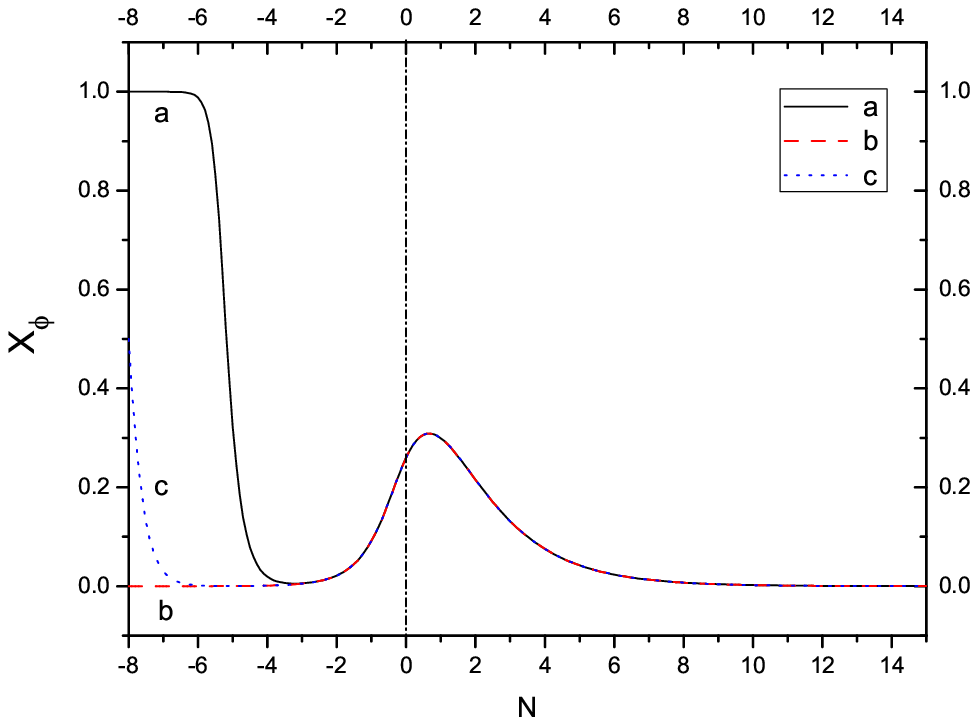}
\includegraphics[height=2.5in,width=3.2in]{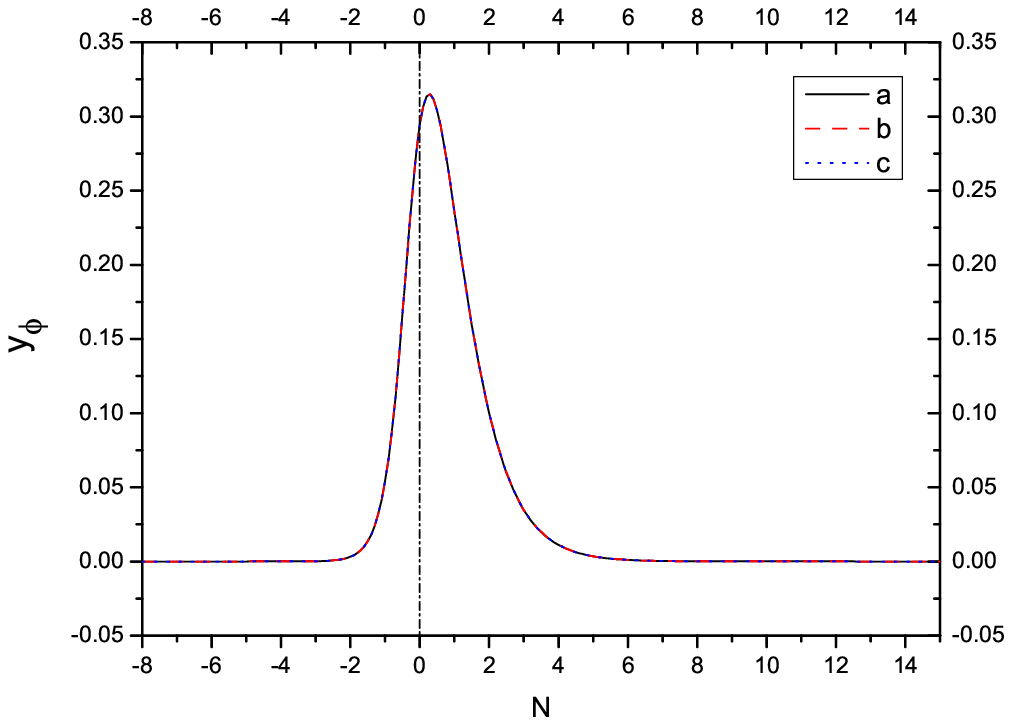}
\includegraphics[height=2.5in,width=3.2in]{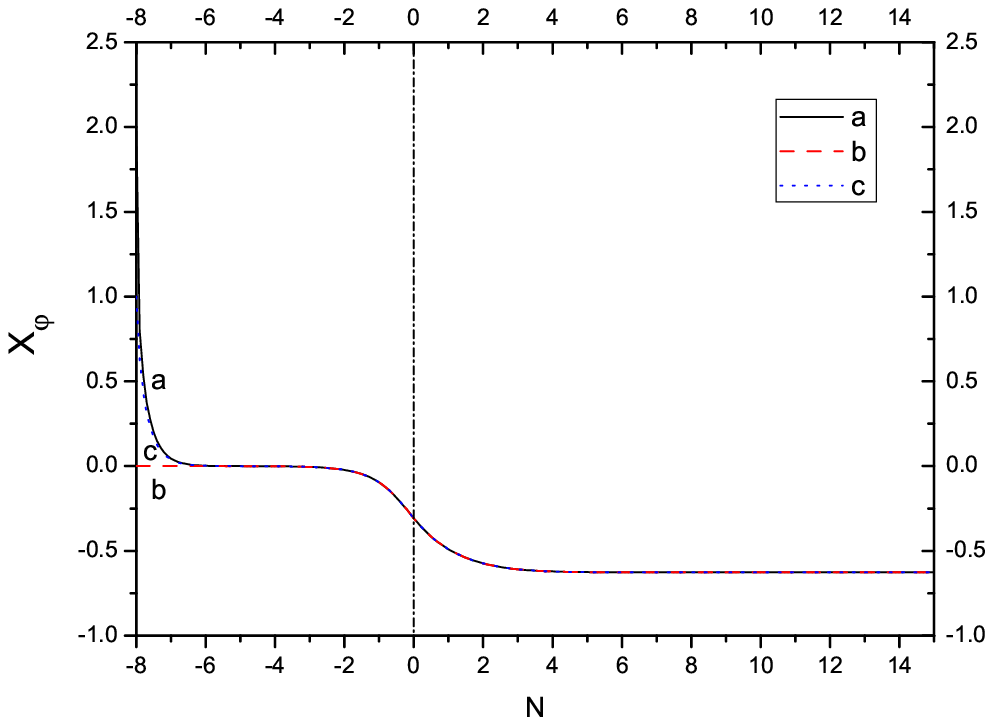}
\includegraphics[height=2.5in,width=3.2in]{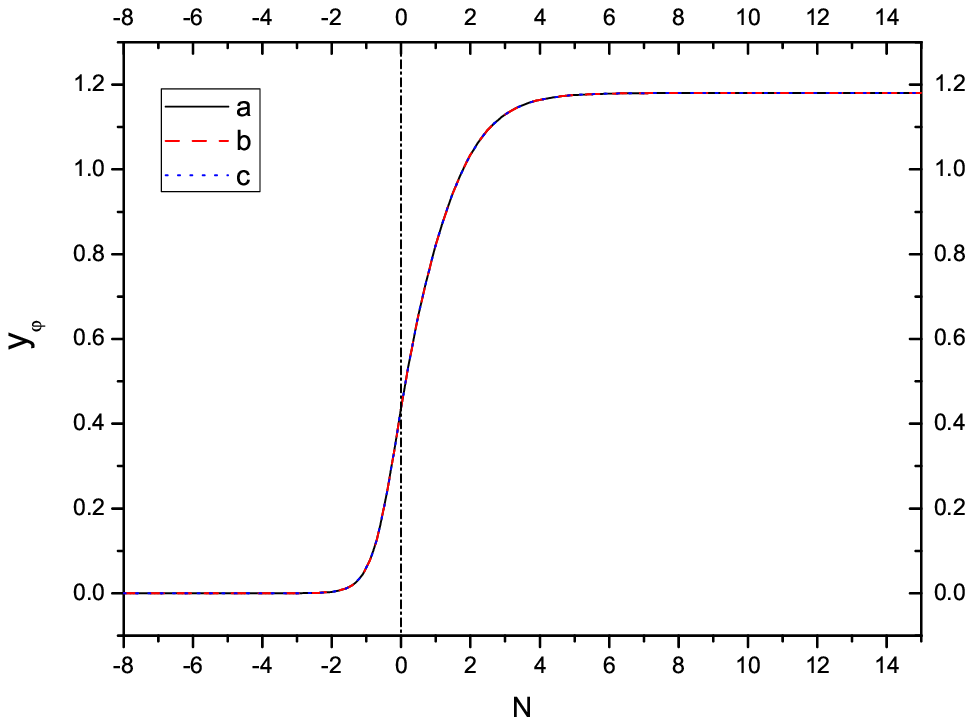}
\caption{\label{figure2} Evolution of the~
$x_\phi,~y_\phi,~x_\varphi$~and ~$y_\varphi$~ as a function of N for
the dark energy model with
$\beta_\phi=\beta_\varphi=1/3,~\gamma=1$.~Initial conditions~(at
$N=-8$) :~~a.~solid line:~$x_{\phi
 i}=0.9999999$,~$y_{\phi i}=6\times10^{-11}$,~$x_{\varphi i}=2.0$,~$y_{\varphi i}=6.5\times10^{-11}$;~~b.~dashed
line:~$x_{\phi i}=1\times10^{-12}$,~$y_{\phi
i}=6\times10^{-11}$,~$x_{\varphi i}=1\times10^{-12}$,~$y_{\varphi
i}=6.5\times10^{-11}$;~~c.~dotted line:~$x_{\phi i}=0.5$,~$y_{\phi
i}=6\times10^{-11}$,~$x_{\varphi i}=1.0$,~$y_{\varphi
i}=6.5\times10^{-11}$.}
\end{figure}
%%%%%%%%%%
From the Fig.\,\ref{figure2}, we can see that the evolution of the~
$x_\phi,~y_\phi,~x_\varphi$~and ~$y_\varphi$~ as a function of N. We
observe that the $y_\phi$ and $y_\varphi$ evolutionary tracks are
not sensitive to the initial conditions of $x_\phi$ and $x_\varphi$
which given in Fig.\,\ref{figure1}. This can be seen from the
Eqs.~(\ref{yundong2}) and (\ref{yundong4}). The initial value of
$x_\phi,~y_\phi,~x_\varphi$~and ~$y_\varphi$ are so small that we
can safely neglect the first three term of the right hand side of
Eqs.~(\ref{yundong2}) and (\ref{yundong4}). So when $-8<N<-2$, we
get $y'_\phi\approx3 \gamma y_\phi,~y'_\varphi\approx3 \gamma
y_\varphi$. When $N\geq-2$ , the different initial value of $x_\phi$
and $x_\varphi$ have converged to a common evolutionary track, so
from then on even the value of $x_\phi,~y_\phi,~x_\varphi$~and
~$y_\varphi$ become large, the convergence of $y_\phi$ and
$y_\varphi$ with different initial value of $x_\phi$ and $x_\varphi$
is no change.

Now, consider the case (a) in which the initial value of $x_{\phi
a}$ is large, such as the solid  line in Fig.\,\ref{figure2}~. Since
$x_{\phi a}>\sqrt{\beta_\phi y_{\phi a}}$ , so  $x_{\phi a}$  will
decrease until $x_{\phi a}=\sqrt{\beta_\phi y_{\phi a}}\equiv x_A$
(see Eq.~(\ref{yundong1})).  Next, consider the case (b) in which
the initial value of $x_{\phi b}$  is small, such as the dashed line
in Fig.\,\ref{figure2} . Since $x_{\phi b}<\sqrt{\beta_\phi y_{\phi
b}}$ , so $x_{\phi b}$ will increase until $x_{\phi
b}=\sqrt{\beta_\phi y_{\phi b}}\equiv x_B$ . We have observed above
that the $y_\phi$ evolutionary tracks is not sensitive to the
initial value of $x_\phi$ in these two case. But this doesn't mean
that $x_A= x_B$ , since the corresponding N might be different, and
$y_\phi$ vary with N . In order to show  these two case will
converge, we suppose that
 $x_{\phi a}=x_{\phi b}+\delta$ at certain point N (e.g. $N=-4$). Since $x_{a \phi},~x_{\phi b}$ are very small, and in some sense
 $y_\phi$~ evolutionary track is independent, so the
Eq.~(\ref{yundong1}) can be expanded to lowest order in $\delta$ :
 $\delta'=-3\delta$ . The solution of this equation is  $\delta\propto
e^{-3N}$ .This means that  $\delta$  dacays exponentially, and the
evolutionary track of different initial value of $x_\phi$  given in
Fig.\,\ref{figure2} converge.

Similar analysis can apply to $x_\varphi$. So we have proved that
the evolutionary tracks of the~ $x_\phi,~y_\phi,~x_\varphi$~and
~$y_\varphi$~ of different initial value given in
Fig.\,\ref{figure1} are converge.

\subsection*{C.~The future of the universe}

The critical points correspond to the fixed points where
$x_{\phi}'=0$, $y_{\phi}'=0$, $x_{\varphi}'=0$, $y_{\varphi}'=0$,
which have been calculated and given in Table I, and there are
self-similar solutions with
\begin{equation}
\label{Hubbleeq2}
  \frac{\dot{H}}{H^2}=-\frac{3}{2}\left( {-\frac{y_\phi (\gamma-x^2_\phi)}{\sqrt{1-x^2_\phi}}-\frac{y_\varphi (\gamma+x^2_\varphi)}{\sqrt{1+x^2_\varphi}}+\gamma} \right)
\end{equation}
This corresponds to an expanding universe with a scale factor $a(t)$
given by $a\propto t^p$, where
\begin{equation}
\label{Hubbleeq3}
  p=\frac{2}{3\left( {-\frac{y_\phi (\gamma-x^2_\phi)}{\sqrt{1-x^2_\phi}}-\frac{y_\varphi (\gamma+x^2_\varphi)}{\sqrt{1+x^2_\varphi}}+\gamma} \right)}
\end{equation}

We now study the stability around the critical points given in Table
I. Consider small perturbations $\delta x_\phi$, $\delta y_\phi$,
$\delta x_\varphi$, and $\delta x_\varphi$  about the critical
points $(x_{\phi c}, y_{\phi c}, x_{\varphi c}, y_{\varphi c})$:
$x_{\phi c}\rightarrow x_{\phi c}+\delta x_\phi$, $y_{\phi
c}\rightarrow y_{\phi c}+\delta y_\phi$, $x_{\varphi c}\rightarrow
x_{\varphi c}+\delta x_\varphi$, $y_{\varphi c}\rightarrow
y_{\varphi c}+\delta y_\varphi$.

\begingroup

\begin{table*}

\begin{tabular}
{c@{\hspace{0.3 cm}} c@{\hspace{0.8 cm}} c @{\hspace{0.5 cm}}c
@{\hspace{0.5 cm}}c @{\hspace{0.5 cm}}c } \hline Label & $x_{\phi
c}$ & $y_{\phi c}$ & $x_{\varphi c}$ & $y_{\varphi c}$
 & $Existence$\\ \hline
$A.$ & $0$ & 0 & 0& 0  & all $\gamma$\\
$B.$ & $0$
 & 0 & $-\sqrt{\beta_{\varphi }y_{\varphi c}}$  &$ \frac{\sqrt{\beta_\varphi^2+4}+\beta_\varphi}{2}$  &all $\gamma$ \\
 $C.$ & $\pm1$ & 0 & 0 & 0  & all $\gamma$  \\
$D.$ & $\pm1$ & 0 &$-\sqrt{\beta_{\varphi }y_{\varphi c}}$  &$ \frac{\sqrt{\beta_\varphi^2+4}+\beta_\varphi}{2}$  &  all $\gamma$\\
$E.$ & $1$ & $\frac{1}{\beta_\phi}$ & $0$ &
$0$ & $\gamma=1$\\
$F.$ & $-1$ & $\frac{\beta^2_\varphi y^2_{\varphi c}}{\beta_\phi}$ &
$-\sqrt{\beta_{\varphi }y_{\varphi c}}$  &$ \frac{\sqrt{\beta_\varphi^2+4}+\beta_\varphi}{2}$  & $\gamma=1$\\
$G.$&$\sqrt{\gamma}$&$ \frac{\gamma}{\beta_\phi}$&0&0&$\gamma<\frac{1}{2}(\beta_\phi\sqrt{\beta_\phi^2+4}-\beta_\phi^2)$\\
$H.$&$\sqrt{\beta_\phi y_{\phi c}} $&$\frac{\sqrt{\beta^2_\phi+4}-\beta_\phi}{2}$&0&0&all $\gamma$\\
  \\ \hline
\end{tabular}
\caption{The list of the critical points.}
\end{table*}
\endgroup
\begingroup
\begin{table*}
\begin{tabular}{c @{\hspace{1.2 cm}}c @{\hspace{1.2 cm}}c @{\hspace{1.2 cm}}c @{\hspace{1.2 cm}}c@{\hspace{1.2 cm}} c @{\hspace{1.2 cm}}c} \hline
Label & $m_{1}$ & $m_{2}$ & $m_{3}$ & $m_{4}$ & Stability
 &  \\ \hline
$A.$ & $-3$ & $3\gamma$
&$-3$ &$ 3\gamma$ & unstable  & \\
$B.$ & $-3$
 & $-3x^2_{\varphi_c}$ & $-\frac{3}{2}(2+x^2_{\varphi c})$ &$-3(\gamma+x^2_{\varphi c})$& stable &   \\
$C.$ & $6$ & $3\gamma$ & $-3$ &
 $3\gamma$ & unstable &   \\
$D.$ & $6$ & $-3x_{\varphi c}^2$ &  $-\frac{3}{2}(2+x^2_{\varphi c})$ &$-3(\gamma+x^2_{\varphi c})$ & unstable &  \\
$E.$ & $0$ & $-\frac{3}{2}$ &  $-3$ &3& unstable &  \\
$F.$ &$6+6x_{\varphi c}^2$&$\frac{3}{2}x_{\varphi c}^2$&
$-\frac{3}{2}(2+x^2_{\varphi c})$ &$-3(\gamma+x^2_{\varphi c})$
 & unstable &  \\
$G.$ &$a$&$b$& $-3$ &$3\gamma$
 & unstable &  \\
 $H.$ &$-3\gamma+3x^2_{\phi c}$&$-3+\frac{3x^2_{\phi c}}{2}$&
$-3$ &$3x_{\phi c}^2$
 & unstable &  \\ \hline\\
  \multicolumn{6}{l}{where:~~ $a,b=\frac{1}{4\beta_\phi}\left(\beta_\phi(3\gamma-6)\pm3\sqrt{16\beta_\phi\gamma^2\sqrt{1-\gamma}+\beta^2_\phi(4-20\gamma+17\gamma^2)}~\right)$}\\
  \hline
\end{tabular}
\caption{The eigenvalues and stability of the critical points.}
\end{table*}
\endgroup

Substituting into Eqs.~(\ref{yundong1})$-$(\ref{yundong4}), lead to
the first-order differential equations:
%%%%%%
\begin{eqnarray}
\left(
\begin{array}{c}
\delta x_\phi' \\
\delta y_\phi'\\
\delta x_\varphi'\\
\delta y_\varphi'\\
\end{array}
\right) = {\cal M} \left(
\begin{array}{c}
\delta x_\phi \\
\delta y_\phi\\
\delta x_\varphi\\
\delta y_\varphi\\
\end{array}
\right) \ , \label{uvdif}
\end{eqnarray}
%%%%%%%
where ${\cal M}$ is a matrix that depends upon $x_{\phi c}, y_{\phi
c}, x_{\varphi c}$ and $ y_{\varphi c}$.

The general solution for the evolution of linear perturbations can
be written as
%%%%%%
\begin{eqnarray}
\label{perturbation}
\begin{array}{c}
\delta x_\phi= u_{11}~exp~(m_1N)+u_{12}~exp~(m_2N)+u_{13}~exp~(m_3N)+u_{14}~exp~(m_4N)\\
\delta y_\phi=u_{21}~exp~(m_1N)+u_{22}~exp~(m_2N)+u_{23}~exp~(m_3N)+u_{24}~exp~(m_4N)\\
\delta x_\varphi=u_{31}~exp~(m_1N)+u_{32}~exp~(m_2N)+u_{33}~exp~(m_3N)+u_{34}~exp~(m_4N) \\
\delta y_\varphi=u_{41}~exp~(m_1N)+u_{42}~exp~(m_2N)+u_{43}~exp~(m_3N)+u_{44}~exp~(m_4N)\\
\end{array}
\label{uvdif}
\end{eqnarray}
%%%%%%%

%%%%%%%%%%
\begin{figure}
\centering
\includegraphics[height=2.5in,width=3.2in]{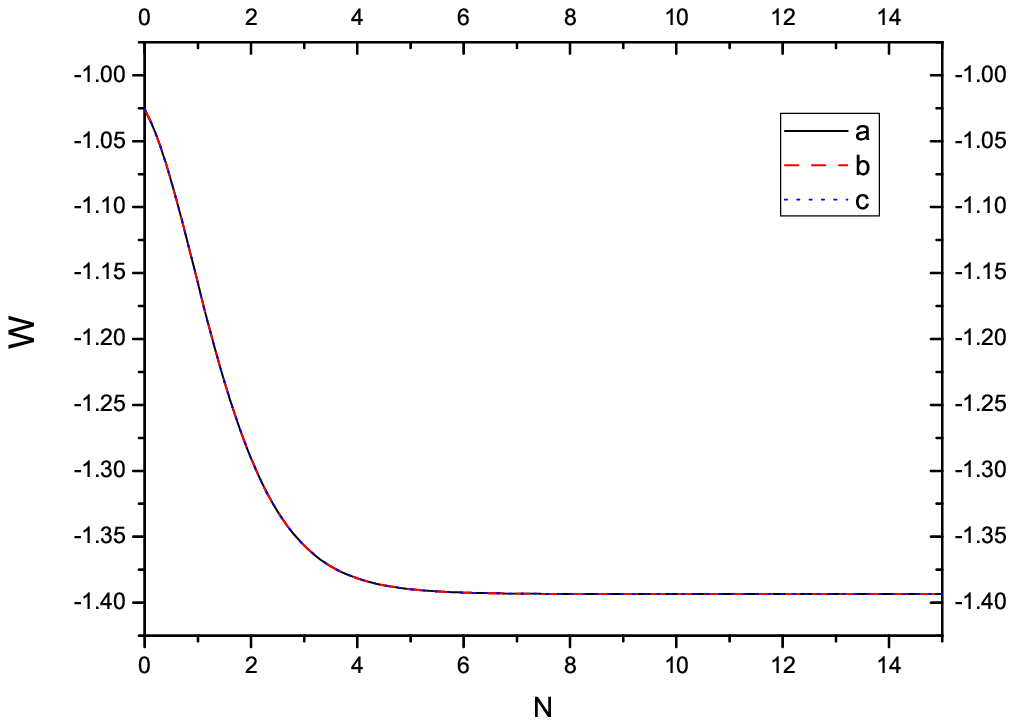}
\includegraphics[height=2.5in,width=3.2in]{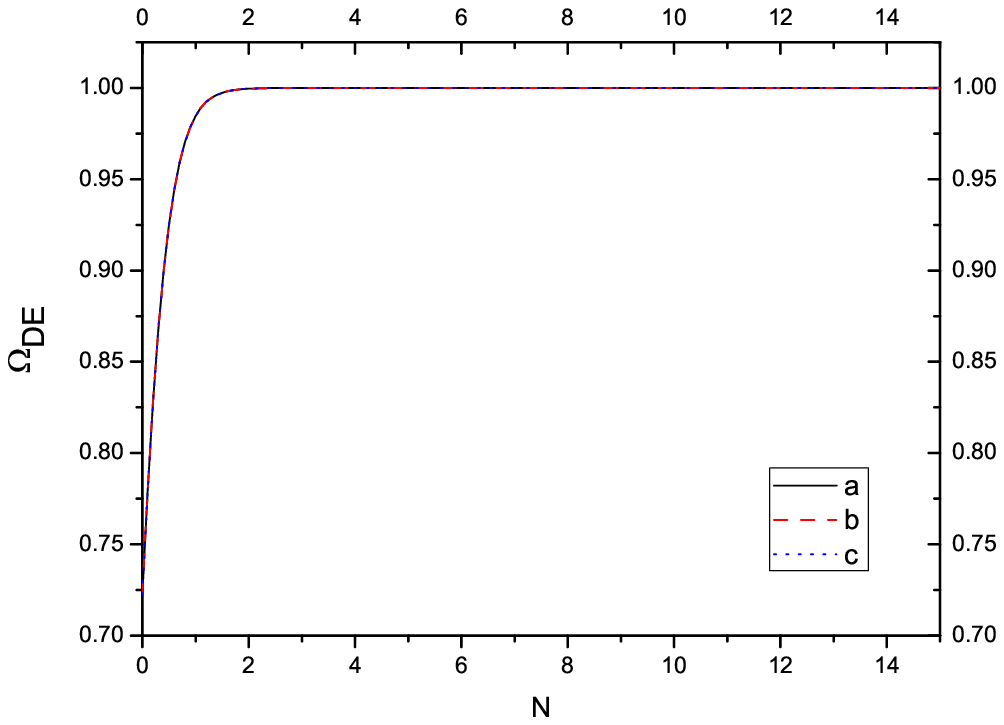}
\caption{\label{figure3} Evolution of the equation of state
($\omega$)and density parameters($\Omega_{DE}$) as a function of N
for the dark energy model with
$\beta_\phi=\beta_\varphi=1/3,~\gamma=1$.~Initial conditions~(at
$N=-8$) :~~a.~solid line:~$x_{\phi i}=0.9999999$,~$y_{\phi
i}=6\times10^{-11}$,~$x_{\varphi i}=2.0$,~$y_{\varphi
i}=6.5\times10^{-11}$;~~b.~dashed line:~$x_{\phi
i}=1\times10^{-12}$,~$y_{\phi i}=6\times10^{-11}$,~$x_{\varphi
i}=1\times10^{-12}$,~$y_{\varphi i}=6.5\times10^{-11}$;~~c.~dotted
line:~$x_{\phi i}=0.5$,~$y_{\phi i}=6\times10^{-11}$,~$x_{\varphi
i}=1.0$,~$y_{\varphi i}=6.5\times10^{-11}$.}
\end{figure}
%%%%%%%%%%
where $m_1$, $m_2$, $m_3$, and $m_4$ are the eigenvalues of the
matrix ${\cal M}$. Thus stability requires the real part of all
eigenvalues being negatives~\cite{CLW97,GPZ02}.

We obtain the eigenvalues and stability for the fixed points in
Table II. The system has a fixed point A which is a fluid-dominated
solution, a fixed point B which is a phantom tachyon-dominated
solution, a fixed point C which is tachyon kinetic-dominated
solution, a fixed point D which is the two-field dominated solution,
two fixed points E and F which exist only for $\gamma=1$, a fixed
point G in which the energy densities $\rho_\phi$ and $\rho_\gamma$
 decrease with the same rate, a fixed H which is tachyon dominated
solution.

In Fig.\,\ref{figure3}, we plot the evolution of the equation of
state in the future. We find that the cosmic evolutionary track is
towards the stable fixed point B in this model. This can be seen by
compare Table I with Fig.\,\ref{figure2}:

Substituting $\beta_\varphi=1/3$ into the stable fixed point B
 ( $x_{\phi c}=0$, $y_{\phi c}=0$, $x_{\varphi
c}=-\sqrt{\beta_{\varphi }y_{\varphi c}}$, $y_{\varphi c}=
(\sqrt{\beta_\varphi^2+4}+\beta_\varphi)/2$ ) , we get $x_{\phi
c}=0$, $y_{\phi c}=0$, $x_{\varphi c}=-0.627285$, $ y_{\varphi
c}=1.180460$, which are consistent with Fig.\,\ref{figure2}. And
from Eq.~(\ref{w}), we know $\omega=-(1+x_{\varphi c}^2)=-1.393487$
at the fix point B, which is consistent with Fig.\,\ref{figure3}.

Generally speaking, if the initial values of
$x_\phi,~y_\phi,~x_\varphi$~and $y_\varphi$ are not the values of
the unstable critical point given in Table I, they will evolve
towards the stable critical point B ( If the physical constraints
$1-x_{\phi }^2>0$ is not violated ). This point can be seen from
Eq.~(\ref{perturbation}). (If one evolutionary track is towards the
unstable critical point, when the values of
$x_\phi,y_\phi,x_\varphi$~and ~$y_\varphi$ differ from $x_{\phi c},
y_{\phi c}, x_{\varphi c}, y_{\varphi c}$ by an amount
$\vec{\delta}$, then from Eq.~(\ref{perturbation}), we know the
$\vec{\delta}$ will be larger, instead of becoming smaller. )

\subsection*{D.~Discussions }
\begin{itemize}
\item
In a tachyon dark energy model, the tachyon is the only source of
the dark energy, and there are three kind stable critical
points~\cite{AL04}, which existence depend on the value of $\gamma$.
But in our model, there is only one stable critical point, which
existence is independent of the value of $\gamma$. In a tachyon dark
energy model, the value of $x_\phi$ and $y_\phi$ can be non-zero at
the stable critical points. But in our model the value of $x_\phi$
and $y_\phi$ must be zero, which we have shown in table I. This's
not accident. This point can be seen as follows. At the critical
point, the value of $x_{\phi c}$, $y_{\phi c}$, $x_{\varphi c}$ and
$y_{\varphi c}$ are fixed, and the value of $x_{\varphi c}$ is
non-zero, otherwise the value of $y_{\varphi c}$ is zero (see
Eq.~(\ref{yundong3})), which means $\rho_{\varphi c}=0$, so
$x_{\varphi c}=0$ is impossible.
\begin{equation}
\label{pw1} \rho_{\phi c}=\frac{V({\phi_c})}{\sqrt{1-\dot{\phi_
c}^2}}=\frac{3M_p^2y_{\phi c}}{\sqrt{1-x^2_{\phi
c}}}H^2~~~~~~~\omega_{\phi c}=\frac{p_{\phi c}}{\rho_{\phi
c}}=-1+\dot{{\phi_c}}^2=-1+x_{\phi c}^2\geq-1
\end{equation}
\begin{equation}
\label{pw2} \rho_{\varphi
c}=\frac{V(\varphi_c)}{\sqrt{1-\dot{\varphi_c}^2}}=\frac{3M_p^2y_{\varphi
c}}{\sqrt{1-x^2_{\varphi c}}}H^2~~~~~~~\omega_{\varphi
c}=\frac{p_{\varphi c}}{\rho_{\varphi c}}=-1-\dot{\varphi_
c}^2=-1-x_{\varphi c}^2<-1
\end{equation}
If $y_{\phi c}\neq0$, then from Eq.~(\ref{pw1}), we know $H^2$ is
nonincreasing at the fix points, since $\rho_{\phi c}$ is
nonincreasing. If $y_{\varphi c}\neq 0$, then from Eq.~(\ref{pw2}),
we know $H^2$ is increasing at the fix points, since $\rho_{\varphi
c}$ is increasing. So either $y_{\phi c}$ or $y_{\varphi c}$ will be
zero . Since $\rho_{\varphi c}$ is increasing and $\rho_{\phi c}$ is
nonincreasing, so we choose $y_{\phi c}=0$. And from
Eq.~(\ref{yundong1}), we know $x_{\phi c}=0$. (This is because
$y_{\phi c}=0$, and if the physical constraints $1-x_{\phi c}^2>0$
is not violated .)

\item
In a tachyon dark energy model with inverse square potential
$V(\phi)=M_\phi^2\phi^{-2}$, in order to have significant
acceleration at late times($a\propto t^p,
p\equiv\frac{1}{2}\left(\frac{M_\phi}{M_p}\right)^2\gg1$), we
clearly require $M_\phi$ much larger than the Planck mass
~\cite{CGST04}. Such a large mass is problematic as we expect
general relativity itself to break down in such a regime. This
problem is fortunately alleviated for the inverse power-law
potential $V=M_\phi^{4-n}\phi^{-n}$ with $0<n<2$. In our model,
since there is another field $\varphi$, with the equation of state
$\omega_\varphi<-1$, a significant acceleration at late times can be
obtained much easier. But the value of $M_\phi$ still can't be small
very much , since if $\beta_\phi=\frac{4M_p^2}{3M_\phi^2}$ is larger
, the risk of $1-\dot{\phi}^2$ becoming nonpositive is increasing,
which can be see from Eqs.~(\ref{yundong1}) and (\ref{yundong2}).

\item
The speed of sound  describe the evolution of small perturbations.
In a tachyon dark energy model the sound speed is
\begin{equation}
c^2_s=\frac{p_{\phi X}}{\rho_{\phi X}}=1-\dot{\phi}^2
\end{equation}
where $X$ denotes the partial derivative with respect to
$X=\frac{1}{2}(\partial_\mu\phi)^2$. Since the value of
$1-\dot{\phi}^2$ is necessarily nonpositive because of the square
root in the Lagrangian density Eq.~(\ref{Lagrangian density1}), the
energy and pressure are real, and inhomogeneous perturbations have a
positive sound speed, so the theory is stable. In our model, there
are two fields. Physically we can use the independent sound speed of
each component to describe the whole system. However, the present
constraints on the sound speed of dark energy are so weak that
considering to study the two independent sound speed is not
justified at present. So it can use the effective sound speed, as
some authors do~\cite{KS84,XCQZZ07}. As we have shown above, when N
is large enough, the fractional energy density of phantom tachyon
$\Omega_\varphi\rightarrow 1$. So ultimately the effective
Lagrangian density is Eq.~(\ref{Lagrangian density2}), and the
effective sound speed is
\begin{equation}
c^2_s=\frac{p_{\varphi X}}{\rho_{\varphi X}}=1+\dot{\varphi}^2>1
\end{equation}
This means that perturbations of the background scalar field can
travel faster than light as measured in the preferred frame where
the background field is homogeneous. But there is no violation of
causality. The theory of the k-essence-like scalar fields with the
Lorentz invariant action is not possible create closed time-like
curves in the Friedmann universe and hence we cannot send the signal
to our own past using the superluminal signals build out of the
¡°superluminal¡± scalar field perturbations~\cite{ECSPM01}.

\end{itemize}

\section{ANOTHER TACHYON-QUINTOM MODEL INCLUDING THE INTERACTION BETWEEN TWO FIELDS }
In order to show some impact of interactions between the two
 scalars on the evolution of the universe, we consider another system, which
include a fluid with barotropic equation of state
$p_\gamma=(\gamma-1)\rho_\gamma$, $0<\gamma\leq2$, and two scalars
with interaction between them. Maybe there are many more
interactions between the two scalars, but for simplicity we only
consider the below Lagrangian density, to see whether there are some
interesting results or not.

The Lagrangian density of the scalars we choose are:
\begin{equation}
\mathcal{L}=-V(\phi,\varphi)\sqrt{1+g^{\mu\nu}\partial_\mu\phi\partial_\nu\phi-g^{\mu\nu}\partial_\mu\varphi\partial_\nu\varphi}
\end{equation}
In this section, we turn our attention to the possibility of the
scalars as a source of the dark energy. We restrict to spatially
homogeneous time dependent solutions for which $\partial_i\phi
=\partial_i\varphi = 0$. Thus the energy densities and the pressure
of the fields are
\begin{equation}\label{eee1}\rho=\frac{V(\phi,\varphi)}{\sqrt{1-\dot{\phi}^2+\dot{\varphi}^2
 }},~~~~p=-V(\phi,\varphi)\sqrt{1-\dot{\phi}^2 +\dot{\varphi}^2}
\end{equation}

Here a dot is derivation with respect to synchronous time. The
background equations of motion are

\begin{equation}
\label{eq11}
\frac{\ddot{\phi}+\ddot{\phi}\dot{\varphi}^2-\ddot{\varphi}\dot{\phi}\dot{\varphi}}{1-\dot{\phi}^2+\dot{\varphi}^2}+3H\dot{\phi}+(1+\dot{\varphi}^2)\frac{dV(\phi)}{Vd\phi}=0
\end{equation}

\begin{equation}\label{eq22}\frac{\ddot{\varphi}-\ddot{\varphi}\dot{\phi}^2+\ddot{\phi}\dot{\phi}\dot{\varphi}}{1-\dot{\phi}^2+\dot{\varphi}^2}+3H\dot{\varphi}-(1-\dot{\phi}^2)\frac{dV(\varphi)}{Vd\varphi}=0
\end{equation}

\begin{equation}\dot{\rho}_\gamma=-3\gamma H\rho_\gamma
\end{equation}

\begin{eqnarray}
\label{Hubble1} \dot{H}=-\frac{1}{2M_p^2} \left(
{\frac{\dot{\phi}^2V(\phi,\varphi)}{\sqrt{1-\dot{\phi}^2+\dot{\varphi}^2}}-\frac{\dot{\varphi}^2V(\phi,\varphi)}{\sqrt{1-\dot{\phi}^2+\dot{\varphi}^2}}+\gamma\rho_\gamma}
\right)
\end{eqnarray}

together with a constraint equation for the Hubble parameter:
\begin{equation}
\label{Hubbleeq1}
 H^2=\frac{1}{3M_p^2}\left(
{\frac{V(\phi,\varphi)}{\sqrt{1-\dot{\phi}^2+\dot{\varphi}^2}}+
\rho_\gamma} \right)
\end{equation}

The potentials we considered are still inverse square potentials:
\begin{equation}
\label{V1}
 V(\phi,\varphi)=\lambda_1M_p^2\phi^{-1}\varphi^{-1}+\lambda_2M^2_p\phi^{-2}+\lambda_3M_p^2\varphi^{-2}
\end{equation}

We define the following dimensionless quantities :
\begin{equation}
\label{df2}
 x_\phi=\dot{\phi},~y_\phi=\frac{\phi^{-1}}{\sqrt{3}H},~X_\varphi=\dot{\varphi},~y_\varphi=\frac{\varphi^{-1}}{\sqrt{3}H},~z=\frac{\rho_\gamma}{3H^2M_p^2}
\end{equation}
Now the Eqs.~(\ref{Hubbleeq1}) and (\ref{Hubble1}) can be rewrite as
follow:
\begin{equation}
\label{HH2} 1=\frac{\lambda_1y_\phi y_\varphi+\lambda_2y_\phi^2
+\lambda_3 y_\varphi^2}{\sqrt{1-x_\phi^2+x_\varphi^2}}+z=
\Omega_{DE} +z
\end{equation}

\begin{equation}
\frac{H'}{H}=-\frac{3}{2}\left[\frac{-\lambda_1y_\phi
y_\varphi(\gamma-x_\phi^2+x_\varphi^2)-\lambda_2y^2_\phi(\gamma-x_\phi^2+x_\varphi^2)
-\lambda_3y^2_\varphi(\gamma-x_\phi^2+x_\varphi^2)}{\sqrt{1-x_\phi^2+x_\varphi^2}}+\gamma\right]
\end{equation}

where $\Omega_{DE}$ measure the dark energy density as a fraction of
the critical density,~a prime denotes a derivative with respect to
the logarithm of the scale factor, $N={\rm ln}\,a$.

Then the evolution Eqs.~(\ref{eq11})and (\ref{eq22}) can be written
to an autonomous system:

\begin{eqnarray}
\begin{array}{c}
\label{em1}
x'_\phi=-3(1-x_\phi^2)\left[x_\phi+\left(\frac{-\sqrt{3}\lambda_1y_\phi-2\sqrt{3}\lambda_2y_\phi^2/y_\varphi}{3\lambda_1+3\lambda_2y_\phi/y_\varphi+3\lambda_3y_\varphi/y_\phi}\right)(1+x_\varphi^2)\right]
\\-3x_\phi
x_\varphi\left[x_\varphi-\left(\frac{-\sqrt{3}\lambda_1y_\varphi-2\sqrt{3}\lambda_3y_\varphi^2/y_\phi}{3\lambda_1+3\lambda_2y_\phi/y_\varphi+3\lambda_3y_\varphi/y_\phi}\right)(1-x_\phi^2)\right]
\end{array}
\end{eqnarray}

\begin{eqnarray}
\label{em2}
\begin{array}{l}
 y'_\phi=y_\phi \left[\frac{-3\lambda_1y_\phi
y_\varphi(\gamma-x_\phi^2+x_\varphi^2)-3\lambda_2y^2_\phi(\gamma-x_\phi^2+x_\varphi^2)
-3\lambda_3y^2_\varphi(\gamma-x_\phi^2+x_\varphi^2)}{2\sqrt{1-x_\phi^2+x_\varphi^2}}\right]-\sqrt{3}x_\phi
y^2_\phi+\frac{3}{2}\gamma y_\phi
\\
\end{array}
\end{eqnarray}

\begin{eqnarray}
\label{em3}
\begin{array}{c}
x'_\varphi=-3(1+x_\varphi^2)\left[x_\varphi-\left(\frac{-\sqrt{3}\lambda_1y_\varphi-2\sqrt{3}\lambda_3y_\varphi^2/y_\phi}{3\lambda_1+3\lambda_2y_\phi/y_\varphi+3\lambda_3y_\varphi/y_\phi}\right)(1-x_\phi^2)\right]
\\~+3x_\phi
x_\varphi\left[x_\phi+\left(\frac{-\sqrt{3}\lambda_1y_\phi-2\sqrt{3}\lambda_2y_\phi^2/y_\varphi}{3\lambda_1+3\lambda_2y_\phi/y_\varphi+3\lambda_3y_\varphi/y_\phi}\right)(1+x_\varphi^2)\right]
\end{array}
\end{eqnarray}

\begin{eqnarray}
\label{em4}
\begin{array}{l}
y'_\varphi=y_\varphi \left[\frac{-3\lambda_1y_\phi
y_\varphi(\gamma-x_\phi^2+x_\varphi^2)-3\lambda_2y^2_\phi(\gamma-x_\phi^2+x_\varphi^2)
-3\lambda_3y^2_\varphi(\gamma-x_\phi^2+x_\varphi^2)}{2\sqrt{1-x_\phi^2+x_\varphi^2}}\right]-\sqrt{3}x_\varphi
y^2_\varphi+\frac{3}{2}\gamma y_\varphi
\\
\end{array}
\end{eqnarray}

 The equation of state of the dark energy is

\begin{equation}
\label{wwww}
\omega=\frac{p}{\rho}=-1+\dot{\phi}^2-\dot{\varphi}^2=-1+x_\phi^2-x_\varphi^2
\end{equation}

For simplicity, we confine ourselves to the case
$\lambda_1\neq0,~\lambda_2=\lambda_3=0$, then the
Eqs.~(\ref{em1})$-$ (\ref{em4}) can be reduced to
%%%%%%%%%%
\begin{figure}
\centering
\includegraphics[height=2.5in,width=3.2in]{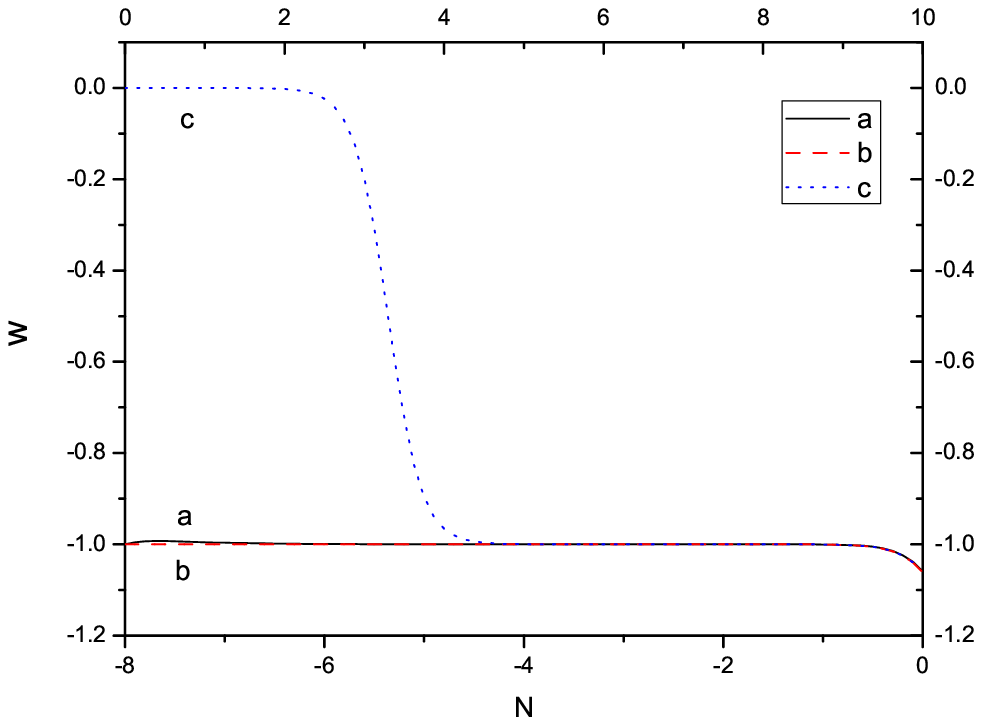}
\includegraphics[height=2.5in,width=3.2in]{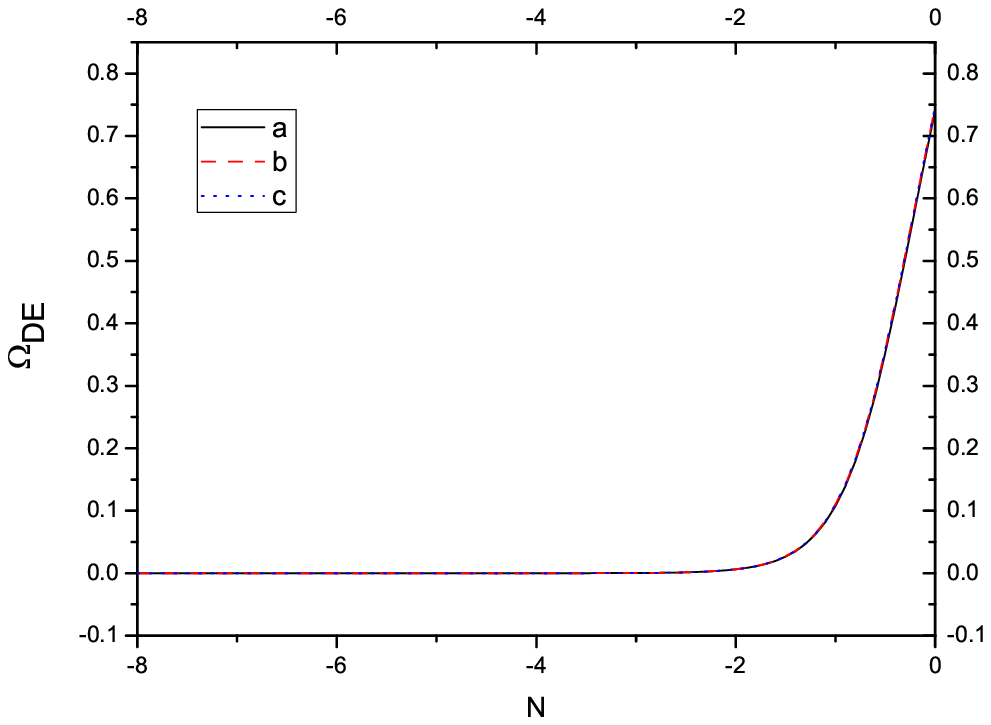}
\caption{\label{figure8} Evolution of the equation of state
($\omega$)and density parameters($\Omega_{DE}$) as a function of N
for the dark energy model with $\lambda_1=1,~\gamma=1$.~Initial
conditions~(at $N=-8$) :~~a.~solid line:~$x_{\phi
i}=1000.0$,~$y_{\phi i}=1\times10^{-5}$,~$x_{\varphi
i}=1000.0$,~$y_{\varphi i}=1\times10^{-5}$;~~b.~dashed
line:~$x_{\phi i}=1\times10^{-12}$,~$y_{\phi
i}=1\times10^{-5}$,~$x_{\varphi i}=1\times10^{-12}$,~$y_{\varphi
i}=1\times10^{-5}$;~~c.~dotted line:~$x_{\phi
i}=0.9999999$,~$y_{\phi i}=1\times10^{-5}$,~$x_{\varphi
i}=1\times10^{-12}$,~$y_{\varphi i}=1\times10^{-5}$.}
\end{figure}
%%%%%%%%%%

\begin{equation}
\label{em5}
x'_\phi=-3(1-x_\phi^2)(x_\phi-\frac{y_\phi}{\sqrt{3}}(1+x_\varphi^2))
-3x_\phi x_\varphi(x_\varphi+\frac{y_\varphi}{\sqrt{3}}(1-x_\phi^2))
\end{equation}

\begin{equation}
\label{em6}
 y'_\phi=\frac{-3\lambda_1y^2_\phi
y_\varphi(\gamma-x_\phi^2+x_\varphi^2)}{2\sqrt{1-x_\phi^2+x_\varphi^2}}-\sqrt{3}x_\phi
y^2_\phi+\frac{3}{2}\gamma y_\phi
\end{equation}

\begin{equation}
\label{em7}
x'_\varphi=-3(1+x_\varphi^2)(x_\varphi+\frac{y_\varphi}{\sqrt{3}}(1-x_\phi^2))
~+3x_\phi x_\varphi(x_\phi-\frac{y_\phi}{\sqrt{3}}(1+x_\varphi^2))
\end{equation}

\begin{equation}
\label{em8}
 y'_\varphi=\frac{-3\lambda_1y_\phi
y^2_\varphi(\gamma-x_\phi^2+x_\varphi^2)}{2\sqrt{1-x_\phi^2+x_\varphi^2}}
-\sqrt{3}x_\varphi y^2_\varphi+\frac{3}{2}\gamma y_\varphi
\end{equation}

The evolution of the equation of state $\omega$ and density
parameters are shown in Fig.\,\ref{figure8} with
$\lambda_1=1,~\gamma=1$. We choose $N=-8$ as the initial number of
e-folds,~so choose the $\gamma=1$ in Eqs.~(\ref{em6}) ~and
(\ref{em8}) is a good approximation. From Fig.\,\ref{figure8} we can
see that this model is not sensitive to the initial kinetic energy
density of the two
fields~($x_\phi=\dot{\phi},x_\varphi=\dot{\varphi}$).
 From Eq.~(\ref{eee1}) and Fig.\,\ref{figure8}, we know that the initial energy density of $\phi$
 varied by about four orders of magnitude is still consistent
with current observational constraints. But the initial potential
energy density of the two fields require fine-tuning to agree with
observations.
\begingroup

\begin{table*}

\begin{tabular}
{c@{\hspace{0.3 cm}} c@{\hspace{0.8 cm}} c @{\hspace{0.5 cm}}c
@{\hspace{0.5 cm}}c @{\hspace{0.5 cm}}c } \hline Label & $x_\phi$ &
$y_\phi$ & $x_\varphi$ & $y_\varphi$
 & $Existence$\\ \hline
$A.$ & $0$ & 0 & 0& 0  & all $\gamma$\\
$B.$ & $\sqrt{\frac{\gamma}{2}}$
 & $ \sqrt{\frac{3\gamma}{2}}$&  0 &0 &all $\gamma$
  \\ \hline
\end{tabular}
\caption{The list of the critical points.}
\end{table*}
\endgroup

\begingroup
\begin{table*}
\begin{tabular}{c c c c@{\hspace{0.8 cm}} c @{\hspace{0.8 cm}}c @{\hspace{0.8 cm}}c} \hline
Label & $m_{1}$ & $m_{2}$ & $m_{3}$ & $m_{4}$ & Stability
 &  \\ \hline
$A.$ & $-3$ & $\frac{3\gamma}{2}$
&$-3$ &$\frac{3\gamma}{2}$ & unstable  & \\
$B.$ & $\frac{1}{2}(-3+3\sqrt{1+2\gamma^2-4\gamma})$
 & $\frac{1}{2}(-3-3\sqrt{1+2\gamma^2-4\gamma})$ & $-3$ &$\frac{3\gamma}{2}$& unstable &
 \\ \hline
\end{tabular}
\caption{The eigenvalues and stability of the critical points.}
\end{table*}
\endgroup

The critical points correspond to the fixed points where
$x_{\phi}'=0$, $y_{\phi}'=0$, $x_{\varphi}'=0$, $y_{\varphi}'=0$,
 which have been calculated and given in Table III. To study
the stability of the critical points, we substitute the linear
perturbations about the critical points into
Eq.~(\ref{em5})$-$~(\ref{em8}) and keep terms to the first-order in
the perturbations. The four perturbation equations give rise to four
eigenvalues. The stability requires the real part of all eigenvalues
be negative (see Table IV for the eigenvalues of perturbation
equations and the stability of critical points).

So there are no stable critical points. Since the two kind critical
points all have $y_\varphi=0$, which means that $z=1$(see
Eq.~(\ref{HH2})). The physical constraints
$1-\dot{\phi}^2+\dot{\varphi}^2>0$  set limit on the equation of
state $\omega$ (see Eq.~(\ref{wwww})) : $\omega<0$. Compare this
with $\gamma$, ($\gamma=4/3$ for radiation, $\gamma=1$ for dust
matter), we know that the energy densities of the fields $\rho$
decrease more slowly than $\rho_\gamma$, so $z=1$ and $y_\varphi=0$
are impossible.

\section{SUMMARY}
In this paper we have studied the tachyon-quintom dark energy
models, in which during the evolution of the universe the equation
of state $w$ changes from $w>-1$ to $w<-1$. Firstly, the model we
studied is made up of two fields, one is tachyon, the other is
phantom tachyon. In order to construct a autonomous system, the
potentials we choose are inverse square potentials. We find the
model is not sensitive to the initial kinetic energy density of
tachyon and phantom tachyon, and we analyze the reason in detail.
The initial energy density (at N=8 ) of the tachyon varied by nearly
four orders of magnitude is still consistent with current
observational constraints. The phase-space analysis of the spatially
flat FRW models shows that there exist a unique stable critical
point, and we compare it with tachyon model at last. Then we
consider another form of two-field model which include the
interaction between two fields. For the case of $\lambda_1\neq0$,~$
\lambda_2=0$,~$\lambda_3=0$, the phase-space analysis shows that
there is no stable critical point. In some sense, this work means
that multiple kessence-like fields can implement the quintom, which
extends the possibilities that the quintom is realized and is worth
further study.

\[ \]
{\bf Acknowledgements:} This work is supported in part by NSFC under
Grant No: 10491306, 10521003, 10775179, 10405029, 10775180, in part
by the Scientific Research Fund of GUCAS(NO.055101BM03), in part by
CAS under Grant No: KJCX3-SYW-N2.

{}

\end{document}